# Indentation of Coatings at micro/nano scale: Crack formation and deflection

A.S.Bhattacharyya

Centre for Nanotechnology
Central University of Jharkhand
Brambe, Ranchi: 835205

Email: 2006asb@gmail.com

**Abstract**
Nanoindentation is an effective way of finding mechanical properties at nanoscale. They are especially useful for thin films where elimination of the substrate effect is required. The mechanism is based upon depth sensing indentation based on Oliver and Pharr modeling. Nanoindentation was used to evaluate the nanomechanical and fractographic properties of hard coatings deposited by magnetron sputtering. Radial and lateral cracks accompanied by delamination of the coatings were observed during the indentation tests. A novel phenomenon of stress induced crystallization was observed in case of the CNx films which resulted in crack deflection. Indentations at micro scale using Vickers Indenter were also done and the cracking phenomena were analyzed.

## 1. The Mechanism of Nanoindentation

Nanoindenting is a new method to characterize material mechanical properties on a very small scale. Features less than 100 nm across, as well as thin films less than 5 nm thick, can be evaluated. Test methods include indentation for comparative and quantitative hardness determination and scratching for evaluation of wear resistance and thin film adhesion. For indentation, the probe is forced into the surface at a selected rate and to a selected maximum force. Nanoindentation studies were performed on the films by Nanoindenter XP (MTS, USA) [1].

Both bulk modulus (E) and hardness (H) are found from nanoindentation. Advantage lies in possibility of very small indentation (depth of the order of 100nm)***.*** Thus it is useful in the case of thin films. During nanoindentation a Berkovich indenter with 70.3˚ effective cone angle pushed into the material and withdrawn. The indentation load and displacement were recorded as shown in *Fig 1* The hardness and elastic modulus are calculated simultaneously. Hardness is defined as the ratio of indentation load and projected contact area (*eqn 1*).

$$H = \frac{P_{\max}}{A_{projected}} \tag{1}$$

where $P_{max}$ is the maximum load applied and $A_{projected}$ is the corresponding projected contact area. The derived hardness is a function of substrate material, thickness of the film and temperature of deposition. Even at small indentation depth there is a strong interaction from the substrate. Thus we have to be very particular in estimating the true hardness of the film without any influence of the substrate. In nanoindentation however it is established that up to 10% of the film thickness there is no substrate effect. The bulk modulus (E) on the other hand is estimated using *eqn 2.*

$$\frac{1}{E_r} = \frac{(1-\nu^2)}{E} + \frac{(1-\nu_i^2)}{E_i} \tag{2}$$

$E_r$ is the reduced modulus and E and v are the elastic modulus and Poisons ration of the test material and $E_i$ and $v_i$ are the elastic modulus ($E_r$) and Poisson's ratio of the indenter material respectively. Since we are using diamond indenter, $E_i$=1141GPa and $v_i$=0.07.

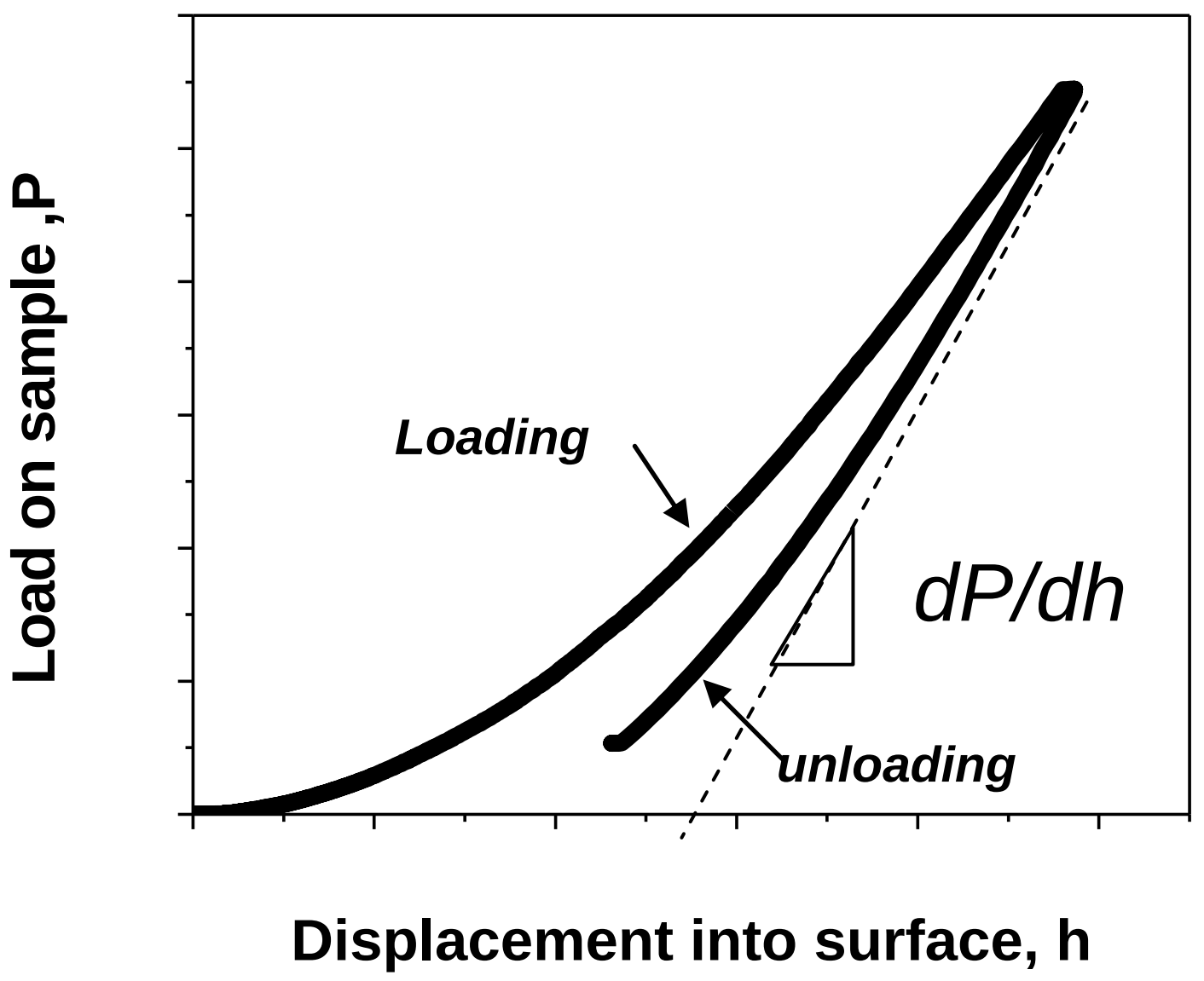

**Fig 1: Load-depth curve during nanoindentation**

The reduced elastic modulus is derived from *dP/dh* is taken from the unloading portion of the curve using *eqn. 3* (*Fig 1*) where *P is the* load on the sample and *h* is the displacement of the indenter and A is the projected area.The projected area is estimated from an empirical function involving the contact depth *($h_c$)* given by *eqn 4.*

$$E_r = \frac{\sqrt{\pi}}{2} \cdot \frac{dP}{dh} \cdot \frac{1}{\sqrt{A}} \qquad (3)$$

The contact depth *($h_c$)* is the depth over which the test material makes contact with the indenter and is determined from Sneddon's equation (*eqn 5)*.

$$A = 24.5 h_c^{\,2} \qquad (4)$$

$$h_c = h_{\max} - \xi \frac{P_{\max}}{S} \qquad (5)$$

where $h_{max}$ is the displacement at peak indentation load $P_{max}$; ξ is a constant depending upon the geometry of the indenter and is 0.75 for Berkovich indenter. *S* is the contact stiffness obtained through curve fitting of the unloading curve of the indentation. The method developed by Oliver and Pharr consists of fitting the load displacement data to the power law equation (*eqn 6*) [1, 2].

$$P = B(h - h_f)^m \tag{6}$$

where P is the load applied to the test surface *h* is the resulting penetration and $h_f$ is the final displacement. B, m are empirically determined fitting parameters. The contact stiffness is then determined by the *eqn 7*

$$S = \left(\frac{dP}{dh}\right)_{h_{max}} = \frac{Bm(h - h_f)^{m-1}}{h_{max}} \tag{7}$$

In order to calculate the hardness (H) and elastic modulus (E) from indentation load displacement data, one must have an accurate measurement of the elastic contact stiffness S and projected area A under load. The Continuous stiffness measurement (CSM) option measures S during loading and not just at the point of initial unload by superimposing a small oscillation on the primary loading signal and analyzing the response by means of an amplifier. This provides hardness and elastic modulus determination as a continuous function of surface penetration. [1- 3].The CSM mode helps in accurately measure the dynamic response of the indenter equipment so as to allow the isolation of material response. It is based on analysis of a simple harmonic oscillator subjected to a forced oscillation. The percentage elastic recoveries of the films were calculated from load depth curve by measuring total area under maximum loading and the retained plastic area using *eqn 3.14* and shown schematically in *Fig 2*.

$$Elastic\ recovery\ \% = \left|\frac{Total\ area - Pastic\ area}{Total\ area}\right| \times 100 \tag{8}$$

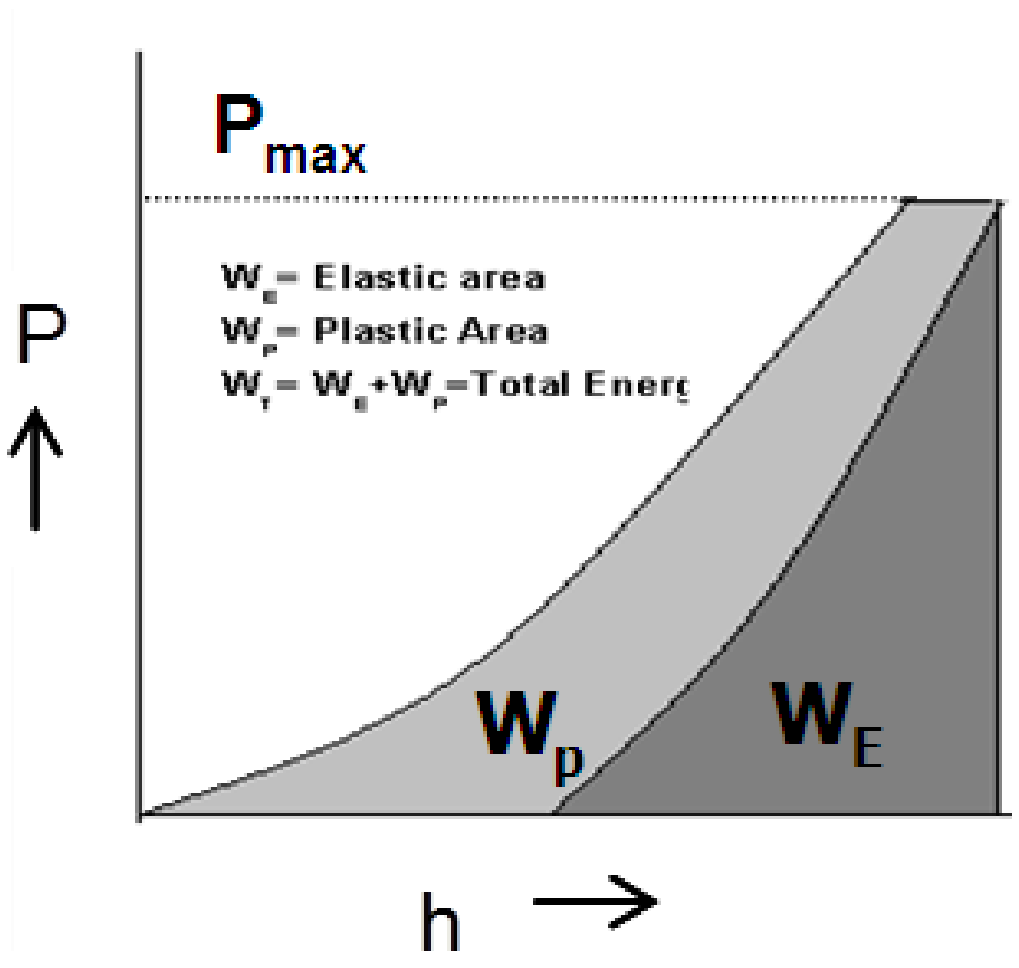

**Fig 2:** The elastic - plastic area in the P-h curve

One of the key elements of nanoindentation is the indenter. The indenter shape has to be precisely calibrated so that correct values of modulus and hardness can be computed. It is desirable to have the indenter material to be both stiffer and harder than the specimen being tested by a large margin. Indenters for nanoindentation are typically constructed from diamond. There are several different types of indenter shapes each having its own advantages for particular applications.

For hard substrates, that 'coating' only properties may be measured where indenter displacements less than 1/10 of the coating thickness *t*. This implies that up to this relative depth both elastic and plastic deformation is concentrated within the coating with the substrate only providing elastic support of a small fraction of the load. A necessary criterion for measuring the plastic properties of the coating is that it yields before the substrate. Thus, the maximum Hertz shear stress not only needs to lie in the coating but also needs to exceed the shear stress of the coating (as the contact load is increased) before the shear stress experienced at depth "*t"* exceeds the shear yield stress of the substrate. For a typical hard coating such as Si-C-N, *H/E* = 0.1 and the critical depth before substrate plastic deformation occurs is 0.126*t*, close to the *t/*10 mentioned above. As we go to softer coatings *H/E* is reduced and the critical *h/t* drops to less than 10%. However, if the substrate is harder than the coating it may not plastically deform in these circumstances and this calculation approach is not valid as plastic deformation extends to the substrate interface in this case and spreads laterally in the soft film. However, this has only a small effect on measured hardness until the constraint of the material trapped between the indenter and the substrate becomes significant. The use of soft substrates (e.g. SS304) will always create the situation where the substrate yields first with a harder coating then being elastically bent and flexed into it followed by coating fracture. Under such conditions, it will be impossible to measure the hardness of the coating itself by indentation tests, at least without changing to a harder substrate. The advantage of using a sharp indenter (Berkovich), plastic deformation starts in the coating at very low loads and the size of the plastic zone increases as the load increases. This zone will grow until its radius is equal to the coating thickness before the substrate beneath it starts to plastically deform [3].

Nanoindentation experiments were done at varying depth of the film with strain rate of 0.05 $s^{-1}$. Poisson's ratio was taken as 0.25 for calculations of all parameters. A harmonic displacement of 2nm at a frequency of 45 Hz was initialized for the experiments. Tip calibration and microscopic calibrations for contact area and positioning respectively were carried out after every 10 indents.

**2. Crack formation during Nanoindentation**

Radial cracks were observed on performing Nanoindentation studies on coatings deposited on silicon substrates. *Fig 3 (a)* shows the formation of radial cracks from the three corners of the indentation impression. The initiation of lateral cracks at higher loads can also be seen as reported in [4]

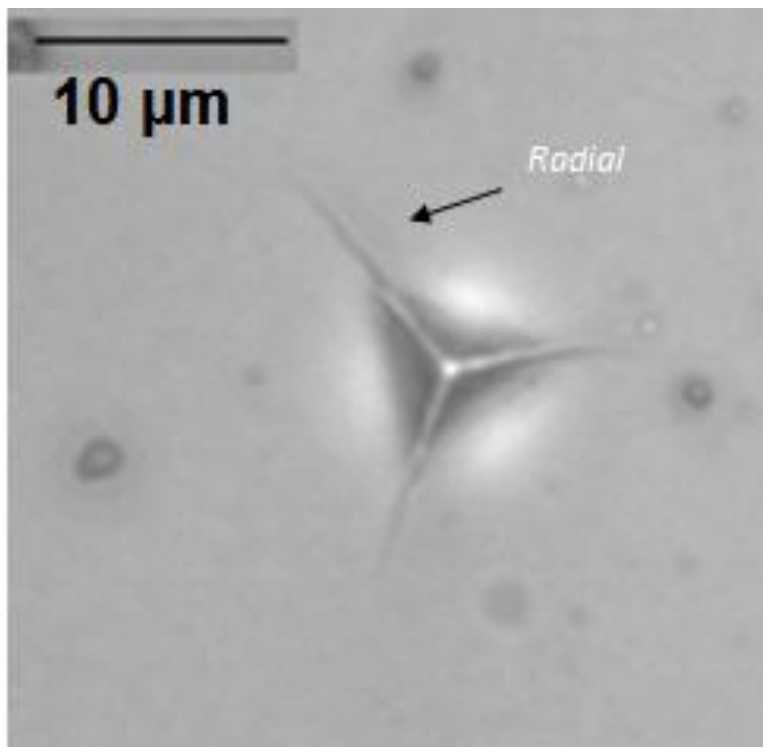

**Fig 3:** Cracks during Nanoindentation of Si-C-N coatings on silicon substrates

### 3. Analysis of *P-h* curve

A nanoindentation Load depth plot for a hard ceramic coating (Si-C-N) is shown in Fig 4. The indentation was done for a depth of 50nm. The plot was fitted with a quadratic curve as shown in the fig 5. The area under the fitted plot was estimated separately to be about 8.2 ($10^{12}$) Joules which is the energy utilized in the indentation process. Fig 6 (a) shows a steady increase in the rate of loading with respect to displacement into surface indicates penetration of the sharp tip efficiently into the surface. The rate of loading with load of the original experimental values is shown in Fig 6 (b).

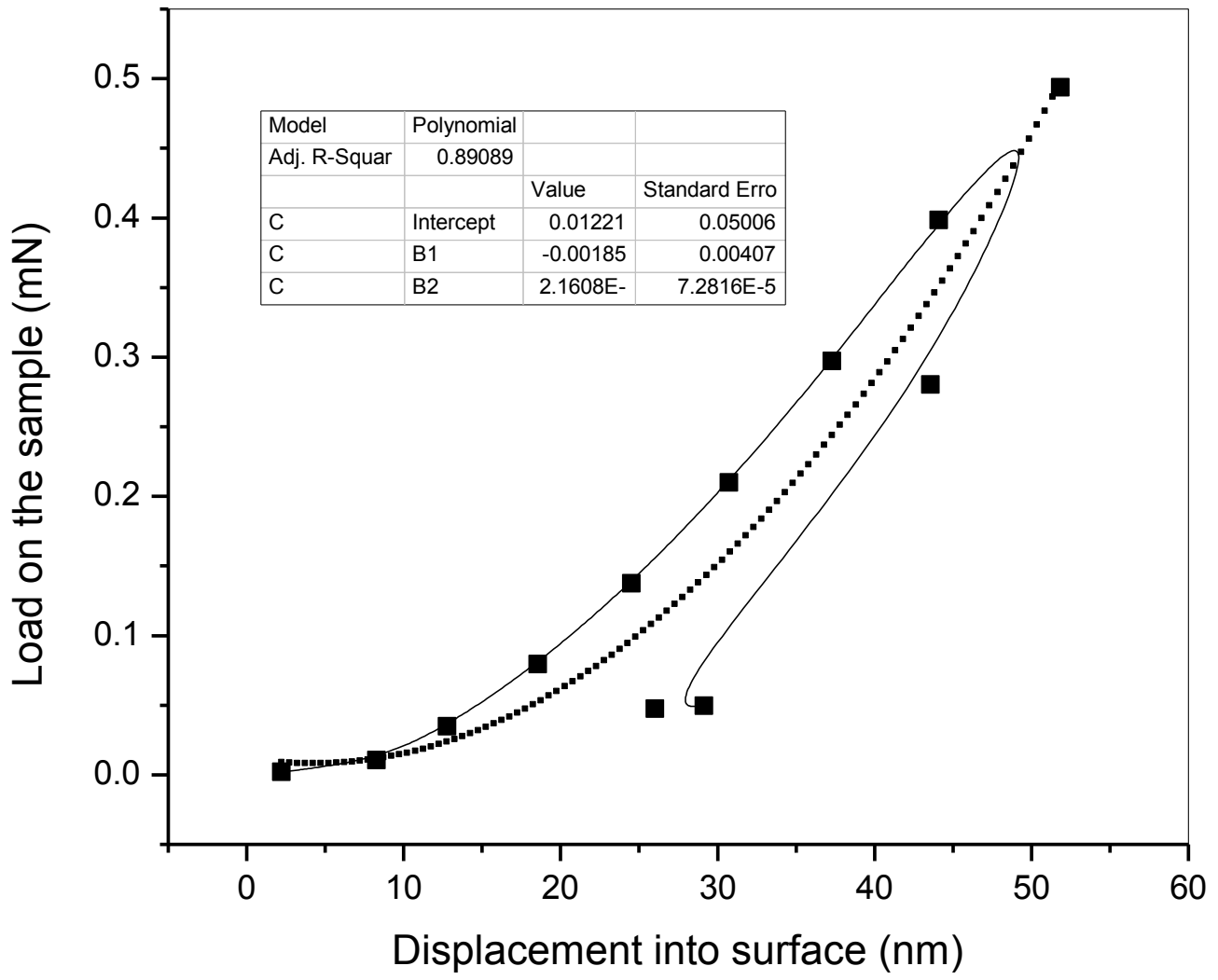

**Fig 4:** Load depth curve during nanoindentation and curve fitting

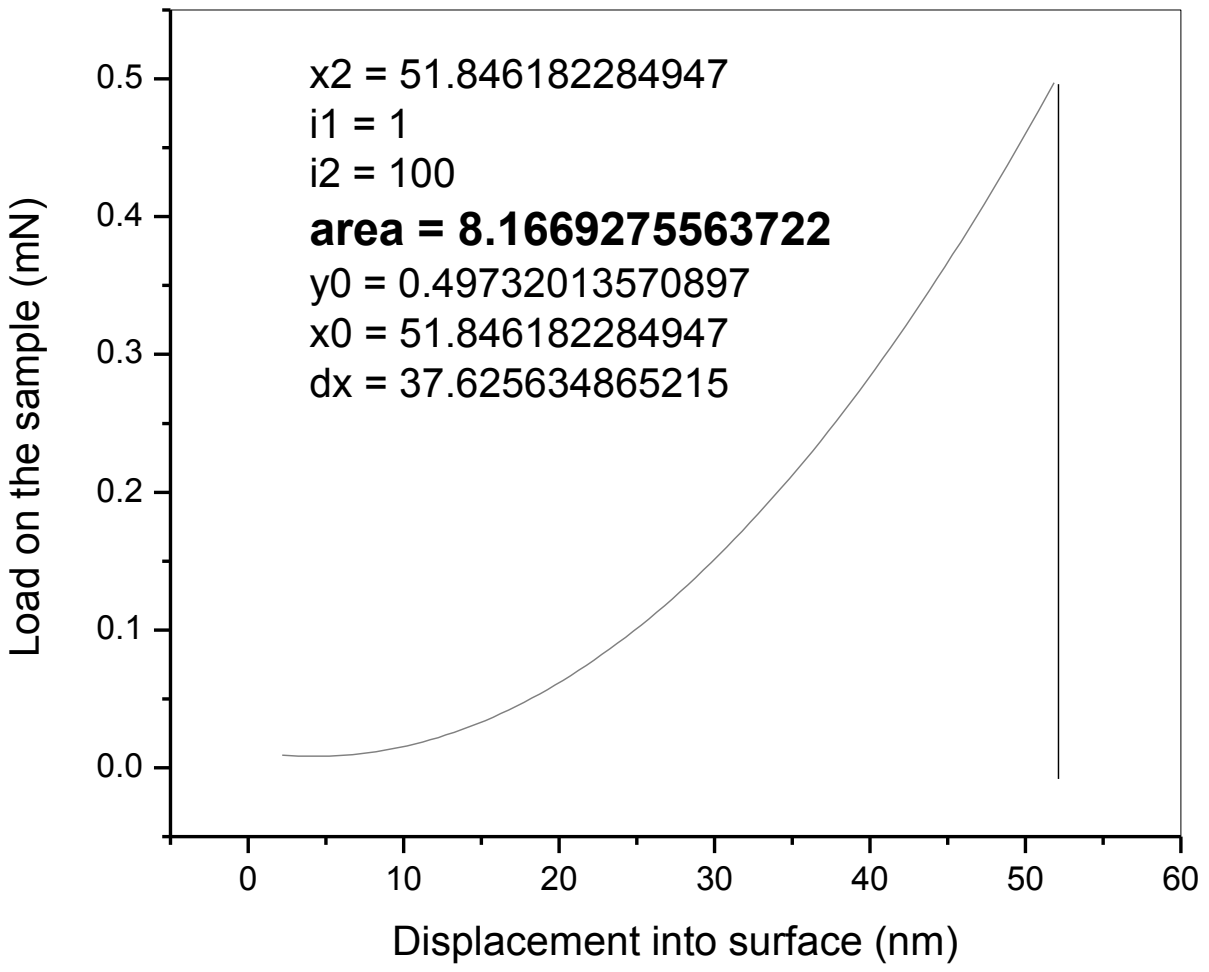

**Fig 5:** Area estimation for the fitted curve

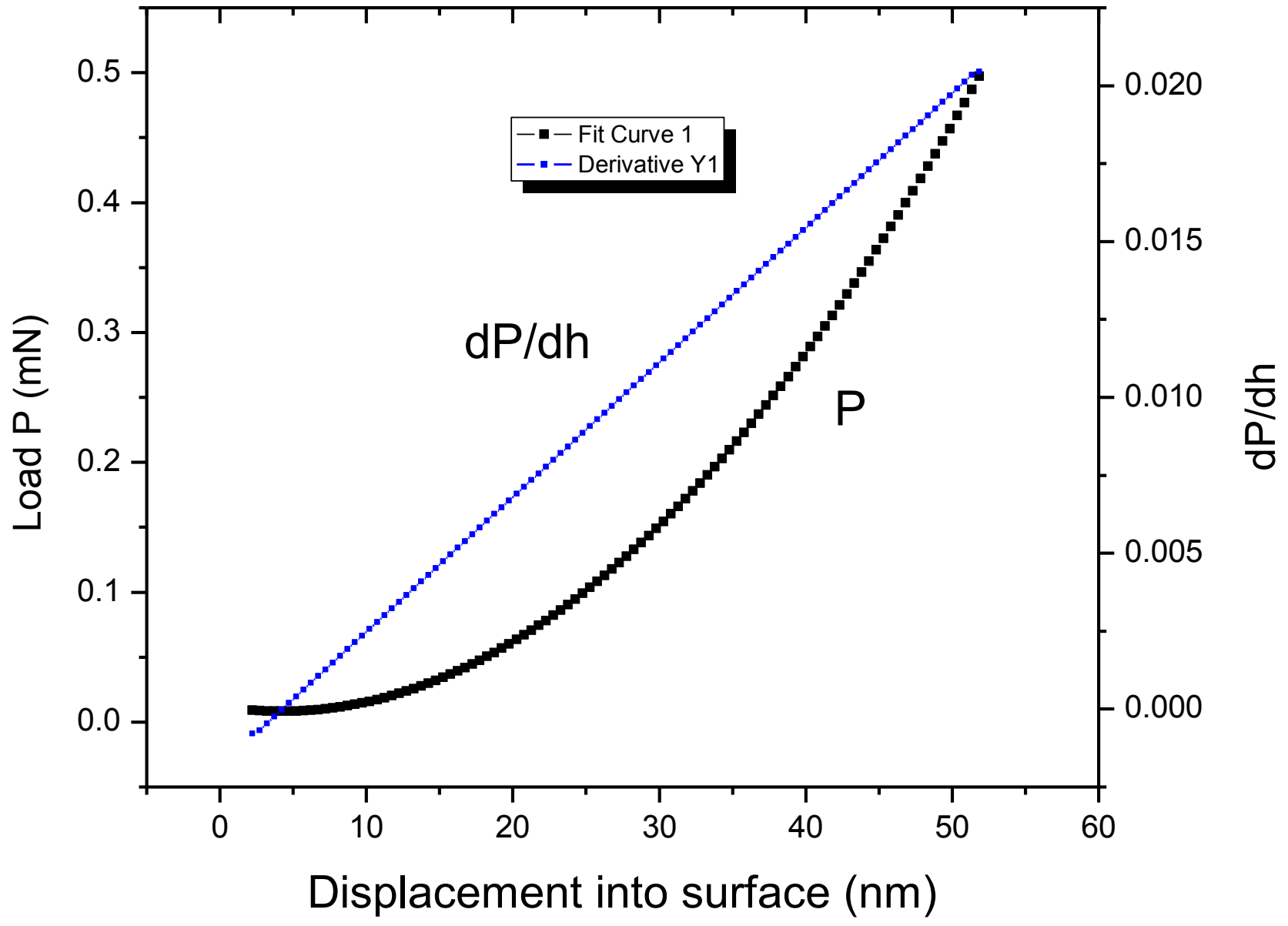

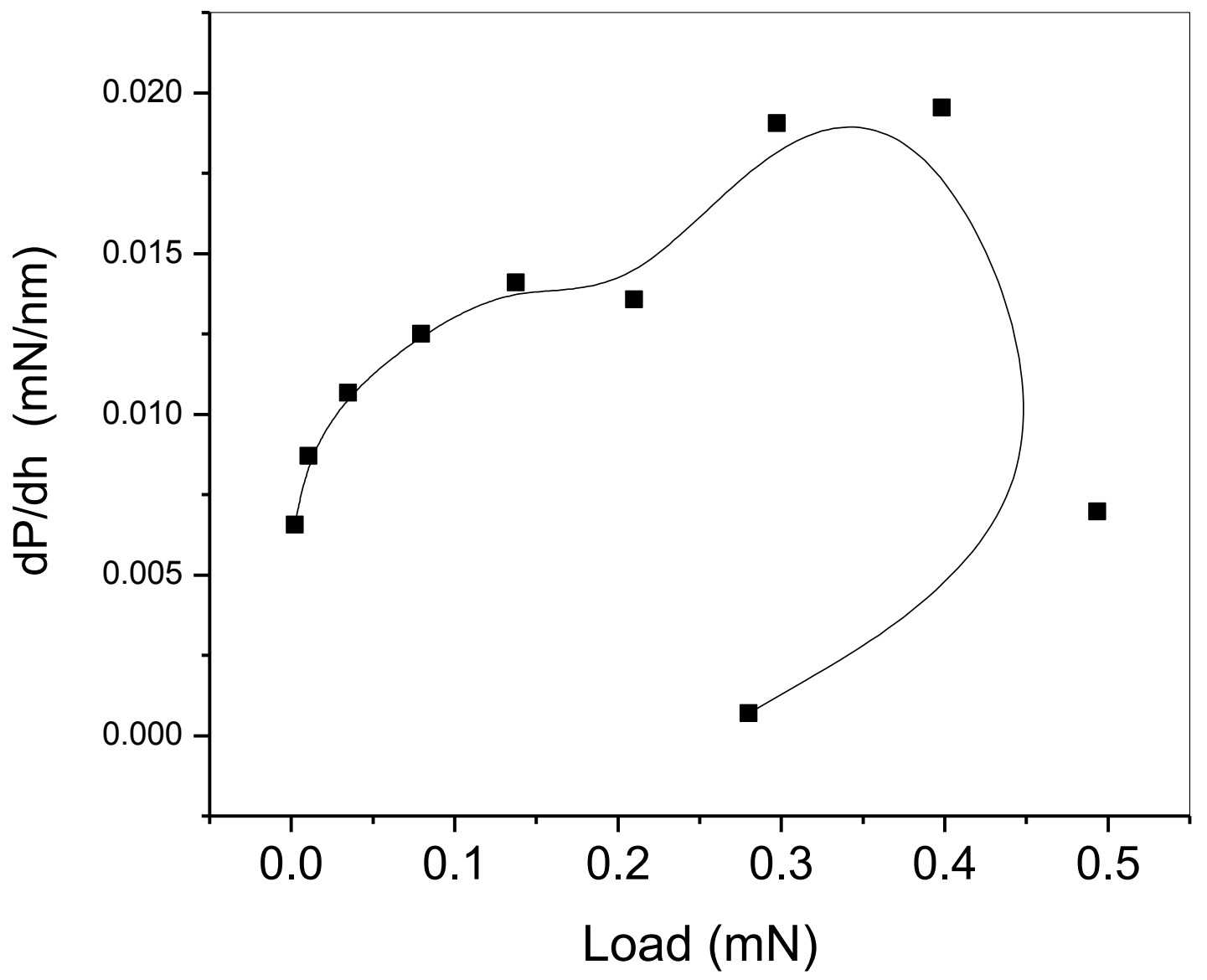

**Fig 6:** The Rate of loading with (a) displacement into surface and (b) loading and unloading loads

## 4. Crack deflection and stress induced crystallization

CNx coatings have been found as efficient wear resistant coatings. CNx coatings have mainly been deposited by various PVD and CVD methods. Nanoindentation of CNx coatings have also been performed and reported in the literature [4, 5]. Nanoindentation is an effective method in determining the nanomechanical properties of the thin films and coatings and is based on depth sensing indentation method. CNx films were deposited by rf magnetron sputtering using a graphite target in Ar/$N_2$ atmosphere. MTS nanoindenter, USA was used for the nanoindentation studies.

In our previous publication we have shown radial and lateral crack formation by nanoindentation on CNx coatings deposited on silicon substrates by radio-frequency (r f) magnetron sputtering [4]. The mechanism of crack formation and deflection by nanoindentation was explained in detail. Crack deflection of small magnitude was observed (Fig 5 a).

This deflection could possibly be due to stress induced crystallization (SIC): a phenomenon which occurs in polymers [6, 7]. However this phenomenon has also been found in systems in non-polymeric materials like Cu-Zr containing bulk amorphous alloys due to deformation-induced nanocrystallites. This phenomenon has also been reported to enhance plasticity in the materials [8]. This portion has been presented in a workshop [9]

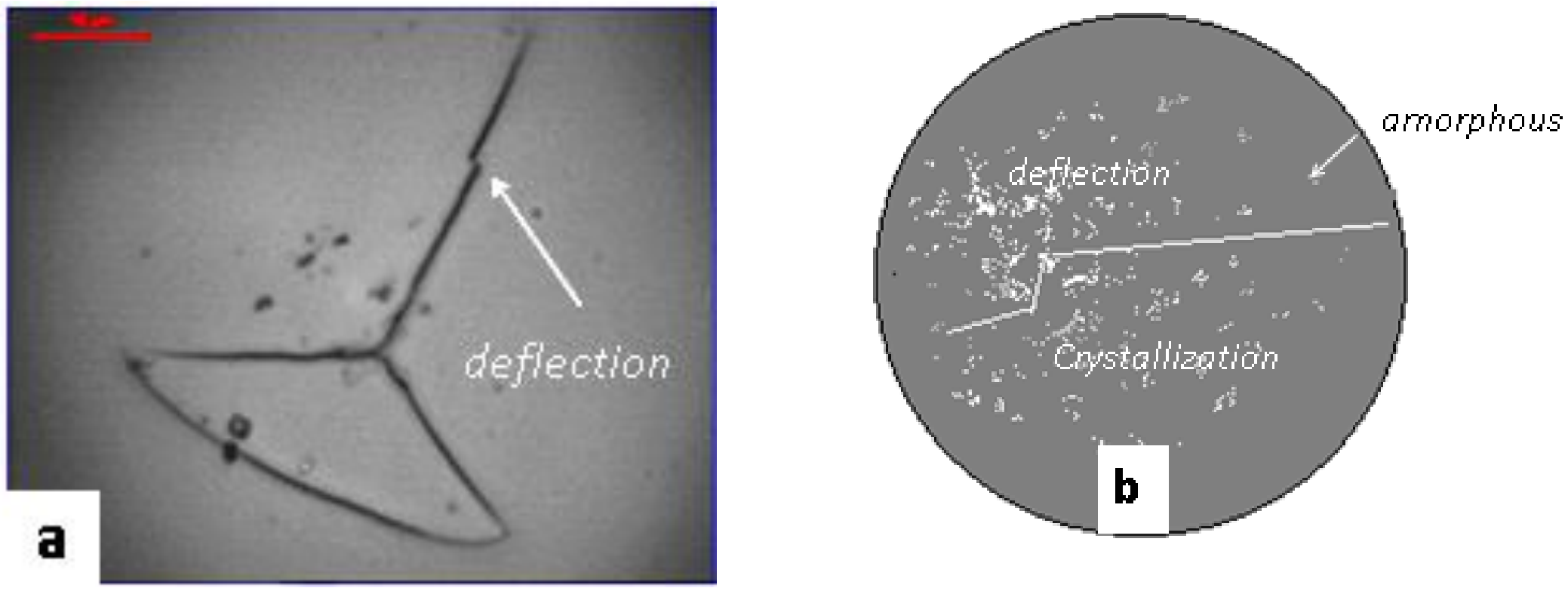

**Fig 7**: (a) Radial and lateral cracks surrounded by circular delamination formed by nanoindentation on CNx coatings deposited by magnetron sputtering [4] and (b) a schematic representation of stress induced crystallization.

## 5. Cracking Phenomenon in Vicker's Microindentation

Similar cracking phenomenon has been observed in case of Vicker's indentation at micro level as well as shown in Fig 6. Similar to microindentaton, radial cracks folloed by chipping and delamination was also observed here. The mechanism of formation of cracks is however slightly different than nanoindentation and also being a microscale deformation, the effect of substartes was dominant here compared to Nanoindentation [10].

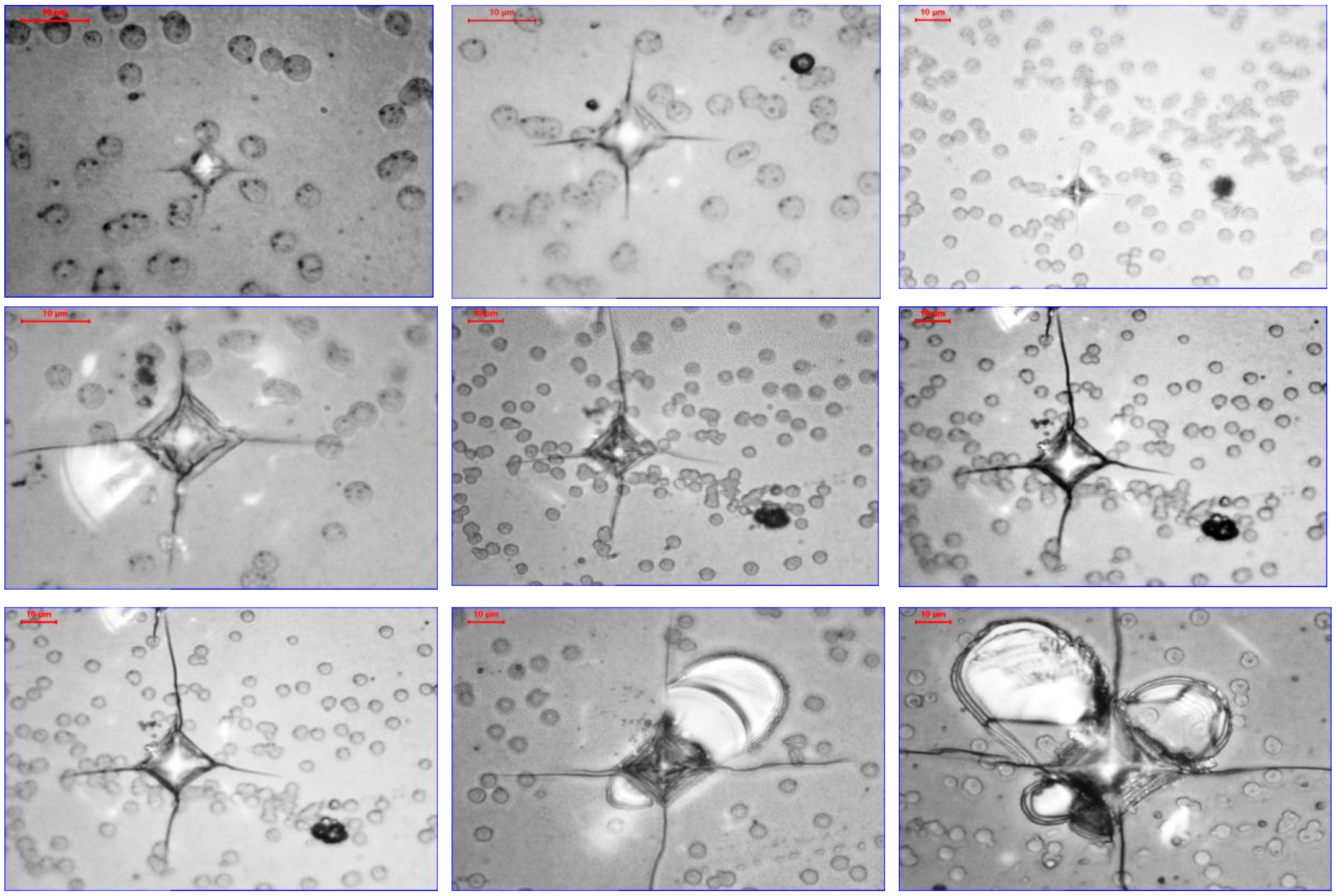

**Fig 8** : Deformation behviour of  nm thick SiCN coatings on silicon

Radial cracks were seen to originate from the corners of the indentation impression. These cracks arise as the elastic free states are not fully recovered. Propagation of theses cracks is prevented until unloading.  Just prior to complete removal of the indenter, they start to grow both in length and depth parallel to the load axis. The maximum depth is of the order of the indentation depth while the crack shape differs from elliptical to circular. Such partially formed cracks tend to propagate downward and merge beneath the plastic zone into half penny geometry as shown in Fig 7[11, 12].

Formations of radial cracks in addition to lateral cracks were observed on increasing the indenter load. These lateral cracks nucleate during unloading and grow outwards from the site of plastic deformation in a plane parallel to the surface (Fig 2). If the applied load is too high for the given specimen then these lateral cracks tend to delaminate the coating by diverting upwards towards the surface and result into material removal from the surface called chipping. The radial cracks are associated with strength degradation whereas the lateral cracks are linked to erosion and wear due to tendency of chipping [12, 13].

The radial crack length (a) was found to increase with increase in load. The variation of $\mathbf{a^{3/2}}$ with load is shown in Fig 7. The coating was 1.5 µm thick. The slope of the curve was used in the Antis formula to determine the fracture toughness of the coating (eqn 9) [14 - 17]. The ratio E/H was found to be 0.1 typical for hard coatings [18].

$$K_C = \beta \cdot \left(\frac{E}{H}\right)^{\frac{1}{2}} \cdot \left(\frac{P_{\max}}{a^{3/2}}\right) \qquad (1)$$

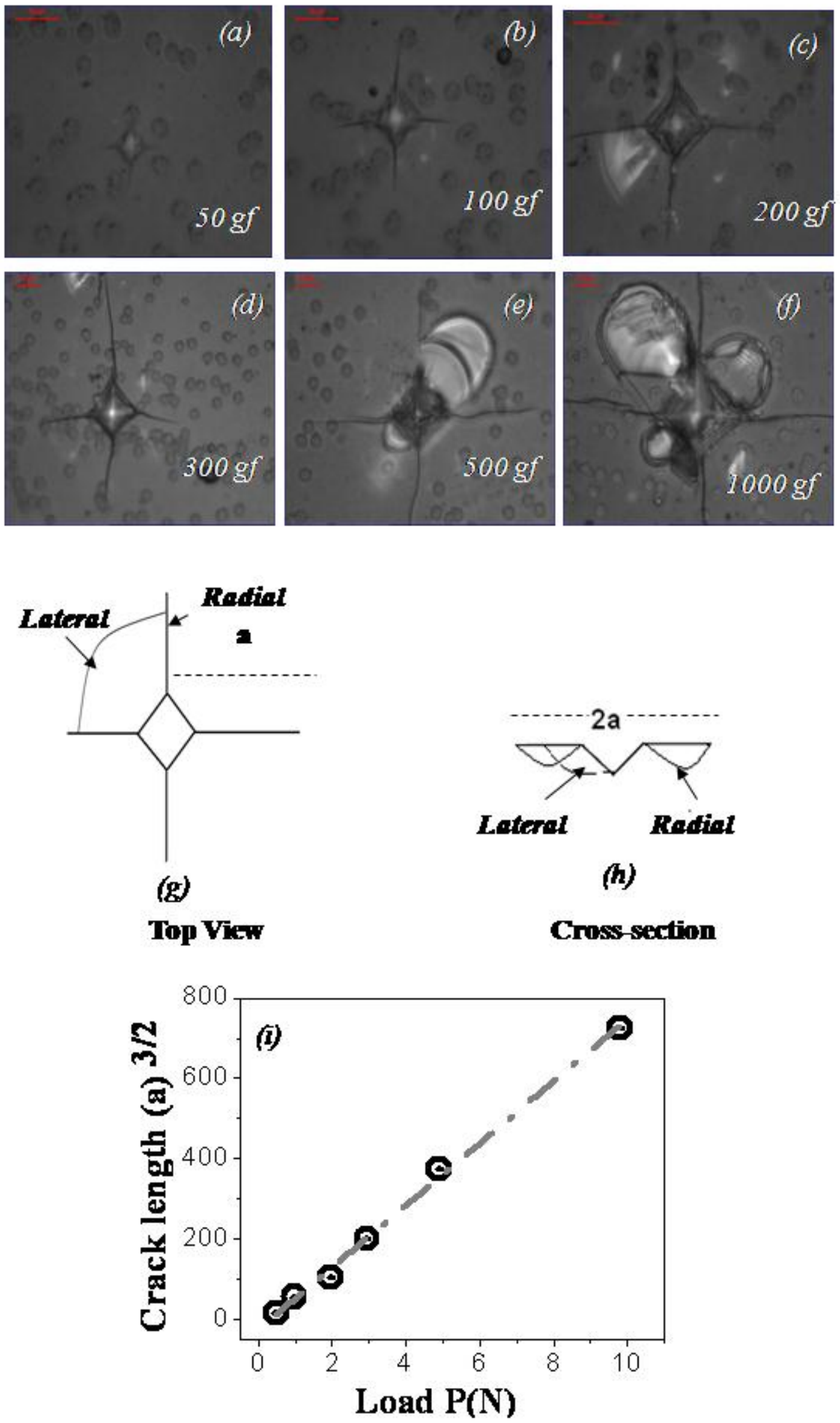

**Fig 9:** Fracture behviour of Si-C-N coatings on silicon substrates at different loads of (a) 50 gf, (b) 100 gf, (c) 200gf, (d) 300gf, (e) 500gf, (f) 1000gf (g) Top view and (h) cross-sectional schematic representation of radial and lateral cracks [12, 18] (i) Graphical evaluation of fracture toughness

The highest toughness of 4.6 MPa√m was observed at a higher thickness of 4.4μm, details of which has been communicated elsewhere [10]. The reason being decreasing substrate influence for increasing coating thickness. Apart from the deceasing substrate influence another parameter that also increases with thickness is the residual stress. At thickness more than 4.4μm the residual stresses developing in the coating becomes large enough to decrease the adhesive strength at the coating substrate interface leading to larger crack propagation and hence reducing the toughness. The thickness at which the highest toughness will be achieved will however vary for other coatings composition. Similarly all the process parameters viz power, pressure, temperature have a role to play in toughness of the Si-C-N coatings, which requires further investigations. For films which are not well adhered to the substarte or which are not having good mechanical strength have shown lower value of toughness as shown in Fig 8

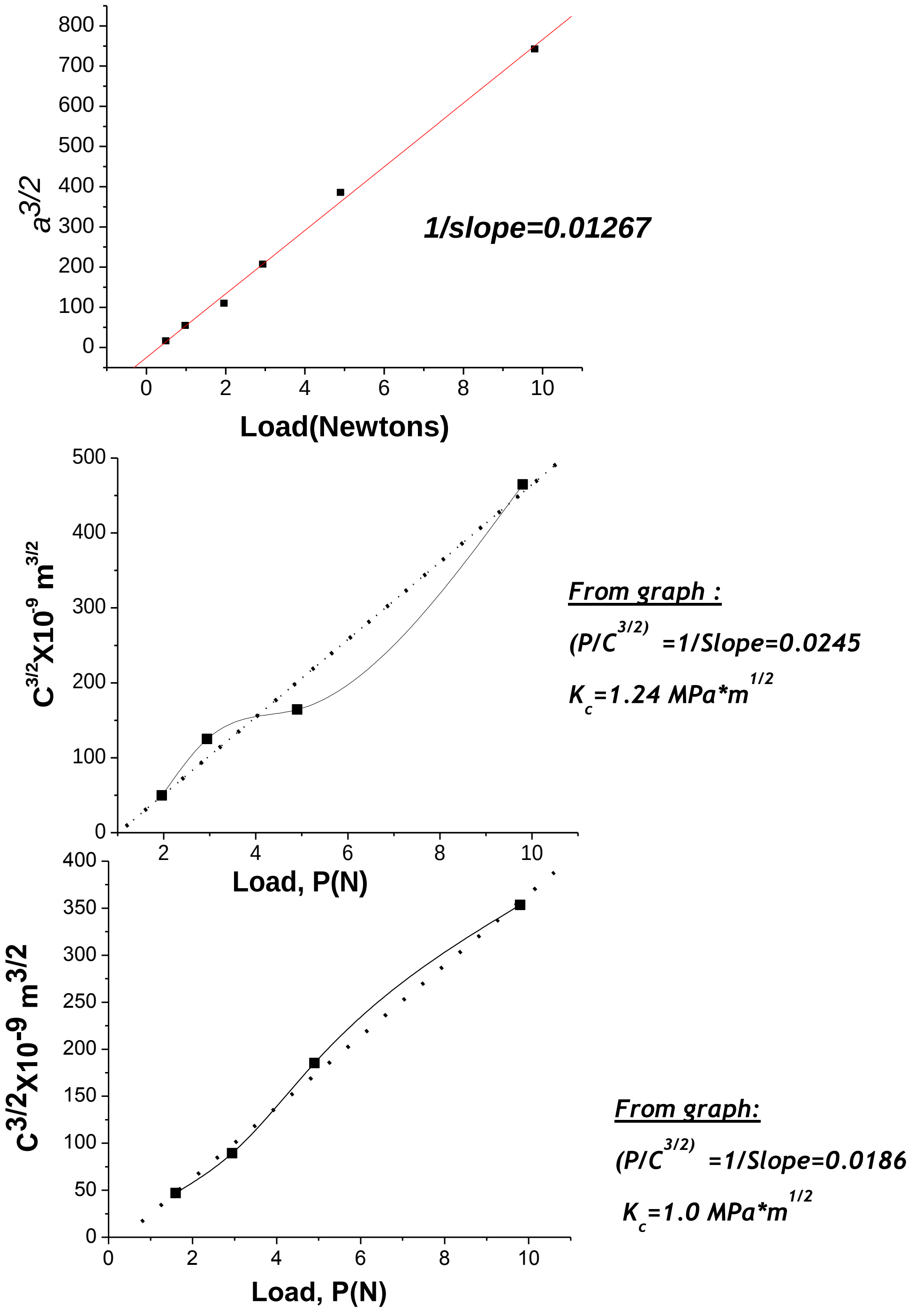

**Fig 10:** Thin film fracture toughness determination graphically from crack length

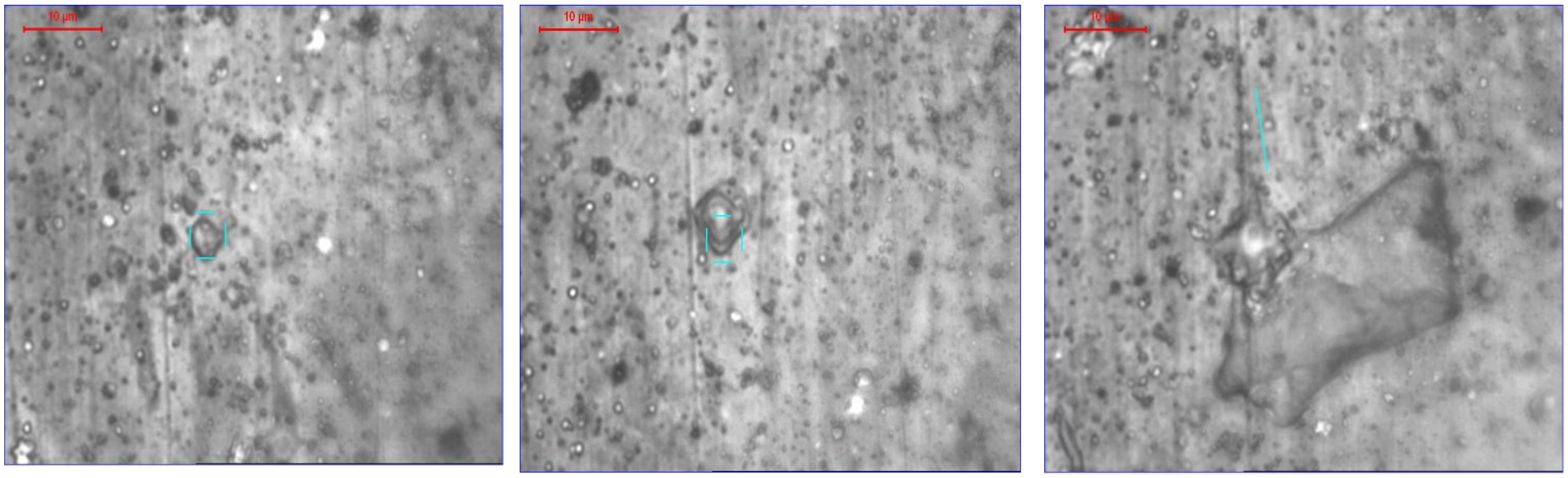

**Fig 11 :** Deformation behaviour of  nm thick  SiCN coatings on 304 SS substrates

For microindentation of hard coating on stainless steel substrates, due to plastic flow of the substrate material, the initial indentations at low loads were circular in nature (Fig 9 a, which were followed by concentric shear band in the indentation impression (Fig 9 b). The radial cracks formed are not that sharp and as in the case of harder substrates the Lateral cracks were also observed but the area was again dominated by plastic deformation (Fig 9 c). A schematic representation of the above process is shown in Fig 10.

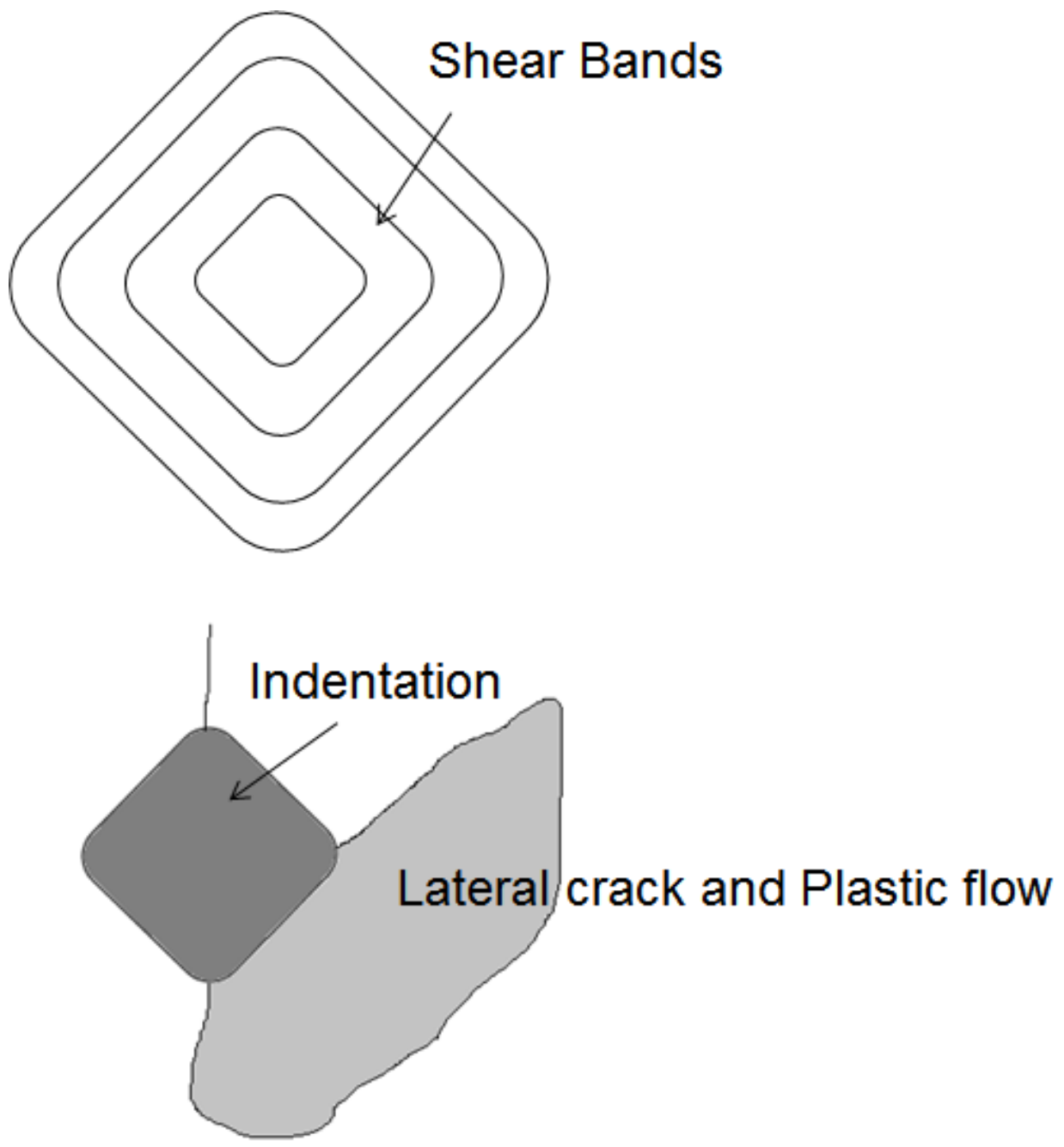

**Fig 12**: Formation of shear bands and plastic flow accompanying Lateral cracks during indentation of hard coating on SS304 substrates

**Acknowledgements**

The author acknowledges Dr. S.K.Mishra, NML Jamshedpur for sputtering and nanoindentation. A part of this work was presented in a PhD thesis [19].